# CONCEPTUALISING REGULATORY CHANGE
## EXPLAINING SHIFTS IN TELECOMMUNICATIONS GOVERNANCE

*Seamus Simpson and Rorden Wilkinson*

*Address for Correspondence:*
Dr Seamus Simpson
Department of Information and Communications
Manchester Metropolitan University
Manton Building, Rosamund Street West
Oxford Road
Manchester, M15 6LL

Tel: 0161 247 3013
Email: S.Simpson@mmu.ac.uk

**Abstract**

Drawing on perspectives from telecommunications policy and neo-Gramscian understandings of international political economy this paper offers an explanation and analysis of the shifting patterns of regulation which have been evident in the telecommunications sector in recent years. It aims to illustrate, explain and explore the implications of the movement of regulatory sovereignty away from the nation-state, through regional conduits, to global organisations in the crystallisation of a world system of telecommunications governance.

Our central argument is that telecommunications governance has evolved from a regulatory arena characterised, in large part, by national diversity, to one wherein a more convergent global multilayered system is emerging. We suggest that the epicentre of this regulatory system is the relatively new World Trade Organisation (WTO). Working in concert with the WTO are existing well-established nodes regulation. In further complement, we see regional regulatory projects, notably the European Union (EU), as important conduits and nodes of regulation in the consolidation of a global regulatory regime.

By way of procedure, we first explore the utility of a neo-Gramscian approach for understanding the development of global regulatory frameworks. Second, we survey something of the recent history – and, in extension, conventional wisdom – of telecommunications regulation at national and regional levels. Third, we demonstrate how a multilayered system of global telecommunications regulation has emerged centred around the regulatory authority of the WTO. Finally, we offer our concluding comments.

**CONCEPTUALISING REGULATORY CHANGE**
EXPLAINING SHIFTS IN TELECOMMUNICATIONS GOVERNANCE

*Seamus Simpson\* and Rorden Wilkinson\*\**

**Introduction**

Key developments in technology and consumption patterns have ensured that telecommunications, more than ever, counts among the most significant sectors of the global economy, having strategic importance for national governments throughout both the industrialised and developing worlds. In terms of manufacturing output, the production of telecommunications equipment to service national and, increasingly, global markets has provided the means to create an infrastructure for users, nurture indigenous manufacturing firms, undertake research and development to stay at the forefront of technological change, improve the balance of trade position of the economy, and provide employment for a significant section of the workforce. Telecommunications service provision has become an increasingly sophisticated, internationally liberalised and hence commercially lucrative activity – a multibillion dollar business, the expansion of which is by no means complete. Indeed, a burgeoning telecommunications sector is regarded as one of the mainstays of healthy service economy, upon which future wealth generation and national economic prosperity is deemed contingent.

Traditionally, the telecommunications sector has been characterised by national, and to some extent regional, variations in regulation, wherein domestic provision and consumption has been subject to the regulatory strictures put in place by state authorities. However, technological advancements and changes in consumption patterns in the sector are now being matched by accompanying developments in the way in which telecommunications is regulated. The extent of these regulatory developments has yet to be fully appreciated, in part the result of their recent occurrence, and, in part because they form one dimension of a wider system of regulation that has traditionally fallen beyond the disciplinary parameters of telecommunications policy. This paper attempts to go someway towards bridging this gap.

---

\* Seamus Simpson is Principal Lecturer in the political and economic aspects of information and communications technologies (ICT) in the Department of Information and Communications, Manchester Metropolitan University. He is author (with Peter Humphreys) of 'European Telecommunications and Globalisation' in Gummett (1996) and has published work on ICT convergence and European Union Information Society Policy.
\*\* Rorden Wilkinson lectures in international political economy in the Centre for International Politics, Department of Government, University of Manchester. He is Convenor of the International Political Economy Group (IPEG) of the British International Studies Association (BISA); author of *Multilateralism and the World Trade Organisation*, (London: Routledge, 2000); and co-editor (with Steve Hughes) of *Global Governance: Critical Perspectives*, (London: Routledge, forthcoming). His work has been published in, among others, the *Journal of World Trade*, *New Political Economy*, *Environmental Politics*, the *British Journal of Politics and International Relations*, *Global Governance*, and *International Studies Perspectives*.

Drawing on perspectives from telecommunications policy and international political economy we offer an explanation and analysis of recent shifts in telecommunications regulation. In doing so, we argue that regional and global organisations are now functioning as key nodes of telecommunications regulation – organisations which together constitute the emergence of a global system of regulation. We argue that this system is eroding national autonomy and diversity in telecommunications provision, replacing it with a more unified and standardised structure reflecting the exigencies of information and communications technologies (ICT) based capitalism. Key to this regulatory convergence is the promotion of a global policy of liberalisation wherein national barriers to market entry are eroded coupled with moves to secure the global ownership of intellectual property and telecommunications rights – developments which have, and will continue, to concentrate ownership in the hands of a small number of global telecommunications players.

To make sense of the growing global convergence around a single model – what Claire Cutler has, in another sense, called a global unification movement (Cutler, 1999) – and central regulative authority, as well as the role of regional and global organisation in this process, we draw on the work of those who have imported something of the work of Italian social theorist Antonio Gramsci to explain events in the global political economy. Against this conceptual backdrop, we examine the processes by which relatively new and still developing regional telecommunications regulatory frameworks are emerging, concentrating our energies on the most advanced example: the European Union (EU). We then explore the relationship between the national, regional and global contexts in a new telecommunications era by considering the role of the World Trade Organisation (WTO) – the emerging epicentre of a wider global economic regulatory framework – in telecommunications regulation. In doing so, we argue that an irreversible movement of regulatory sovereignty away from the nation-state, through regional conduits, to global organisations in the crystallisation of a world system of telecommunications governance is occurring. Such a system, we argue, is primarily geared towards the needs of large-scale transnational telecommunications companies.

By way of procedure we begin by exploring the utility of a neo-Gramscian framework for understanding regulatory change. Second, we survey something of the recent history – and, in extension, conventional wisdom – of telecommunications regulation at the national and regional levels. Third, we demonstrate how a multilayered system of global telecommunications regulation has emerged centred around the growing regulatory authority of the WTO. Finally, we offer our concluding comments.

**Gramsci and regulatory change**

The writings of the inter-war Italian Communist Party leader Antonio Gramsci have inspired a small community of scholars seeking to better understand the complex of economic, ideological, cultural and political factors that together constitute the contemporary world order (see Cox, 1987, 1996; Murphy, 1994; Gill, 1993; Overbeek, 1990; Robinson, 1996; Rupert 1995; Lee, 1996). Like Marx, from whom he drew much (though not all) of his inspiration, Gramsci was concerned with better understanding the relationship between capitalism, its associated social practices and its institutional forms (Rupert, 1995: 16). Gramsci was not, however, concerned with international political economy, let alone global telecommunications regulation. Rather, his aim was to better understand the various factors that enabled Mussolini's fascist regime to establish itself within inter-war Italy, and in doing so explore the potential for building an alternative society. Gramsci's concern with domestic power relations, and the relative lack of comment on the international dimensions of capitalism seem to preclude the use of his work to explore regulatory movements in telecommunications. This is not, however, the case. Through the transposition of certain of Gramsci's concepts to the global level – that is, through a process of internationalisation – we are able to develop alternative ways of thinking about international political economy, and, in extension, telecommunications regulation.

Crudely put, Gramsci's account of the synergistic relationship between capitalism and other dimensions of society led him to develop an understanding of the state that included both the formal trappings of government and the wider realm of social life. For him, it was no longer adequate to talk of the state in a limited governmental sense as the embodiment of capitalist interests. Rather, in order to more fully appreciate the terrain in which capitalist interests were expressed, account also had to be taken of the private realm of individuals: the education system, religious organisations, trade unions, private societies such as the Freemasons and Rotarians, the media, popular culture and the like. Gramsci termed these realms respectively 'political' and 'civil' society. For him, it was only by adopting an expanded notion of the state that a more complete appreciation of the relationship between capitalism and society could be attained. In his refinement of Marx's architectural metaphor, Gramsci describes capitalist society thus:

> What we can do … is to fix two major superstructural 'levels' [upon the economic base of capitalism]: the one that can be called "civil society", that is the ensemble of organisms commonly called "private", and that of "political society" or "the State". These two levels correspond on the one hand to the function of "hegemony" which the dominant group exercises throughout society and on the other hand to that of "direct domination" or command exercised through the State and "juridical" government (Gramsci, 1971: 12).

But this was not the sum of his thinking. Gramsci attributed a particular function to the role of law. For him, law was more than just the means by which order was established; rather, it was one means by which a particular, historically specific social order was established. Law, he argued, functions both actively and passively in shaping the contours of social interaction. Moreover, and in parallel to his expanded conception of the state, Gramsci refused to confine his understanding of law to that associated simply with state judiciary. Instead, he extended it to cover 'the general activity of law'. This, for him, ranged further afield than a simple association with the juridical trappings of government to include those social laws that shape the private life of people, their morality, customs and belief systems (Gramsci, 1971: 195). For Gramsci, then, law in all of its forms constitutes one means by which society could be shaped to meet the changing exigencies of capitalism.

To complete his understanding of capitalist society, Gramsci linked the role of the state and the function of law with its economic base. He deemed the state to be the 'educator' of society and law the means by which social education is undertaken. In tandem, the state and law brought about the adaptation of society to the changing needs of capitalist production. To return to Gramsci:

> In reality, the State must be conceived of as an "educator", in as much as it tends precisely to create a new type or level of civilisation. Because one is acting essentially on economic forces, reorganising and developing the apparatus of economic production, creating a new structure, the conclusion must not be drawn that superstructural factors should be left to themselves, to develop spontaneously, to a haphazard and sporadic germination. The State, in this field, too, is an instrument of "rationalisation", of acceleration and of Taylorisation. It operates according to a plan, urges, incites, solicits, and "punishes"; for, once the conditions are created in which a certain way of life is "possible", then "criminal action or omission" must have a punitive sanction, with moral implications, and not merely be judged generically as "dangerous". The Law is the repressive and negative aspect of the entire positive, civilising activity undertaken by the State (Gramsci, 1971: 247).

At first glance, though interesting, Gramsci's insight into the interrelationship of the constituent elements of all that comprises capitalist society seems to offer little to help conceptualise recent developments in global telecommunications regulation. Part of the problem lies in Gramsci's concentration on understanding the complexities of the state (Kenny and Germain, 1998: 15-16). In the absence of a global state, Gramsci's concepts appear irrelevant. This is not, however, the case. We can take a lead from Gramsci's insistence that international relations naturally developed from national relations, whereupon a reciprocal relationship between the two ensues (Gramsci, 1971: 176). From this we can posit that as capitalist production transcends the boundaries of the state – that is, it begins to be globalised – the jurisdiction of the state internationalises as a reflection of that expansion. The state is internationalised in the sense that its economic base is no longer determined by the exigencies of a nationally-rooted capitalist economy; rather, the determining force of the contemporary state lies with the capitalist world economy. The notion of an internationalisation of the state does not, however, require that the physical boundaries of the state creep beyond national boundaries. Rather, it refers to an extension in the jurisdiction of the state – understood in the Gramsci's sense of political + civil society. More properly, this refers to an internationalisation of political authority (Cox, 1987: 257-8). It follows, then, that the superstructural dimensions of capitalist society will also be reoriented to meet the exigencies of global capitalism. That is, as capitalist production spills beyond the confines of the state, a corresponding political and civil apparatus will develop. Once created, this political and civil apparatus acts to smooth the spread of capitalism. Similarly, law will become internationalised. No longer is the function of law to promote the conformity required of national capitalism, it begins to socialise areas beyond the state through the development of international rules, regulations, norms and codes of behaviour. In this way, we can posit that as production becomes increasingly internationalised, so too does the social, political and legal apparatus essential for smoothing its advance.

We can see something of this internationalisation in Craig Murphy's utilisation of Gramsci's concepts. Murphy argues that one means of better understanding the development of international organisations, the gatekeepers of international economic law, is to explore their relationship with the production process (Murphy, 1994: 9). He argues that international organisations respond to, as well as facilitate changes in, the nature of production. Historically, Murphy suggests, the change from one predominant form of industry has generated the impetus for the development of a particular organisation or set of organisations. In turn, these bodies have smoothed the transition from one form of production to another. The transition from textile production to heavy industry in the mid-nineteenth century, and later, from the chemical, electrical and food-processing industries to car and aircraft production can be seen as having been aided, to some degree, by international organisations. Similarly, international organisations have helped transform the industrial base from manufacturing to information technology and services. As Murphy puts it, 'world organizations have played a role in the periodic replacement of lead industries, a critical dynamic of the world economy since the industrial revolution' (Murphy, 1994: 2).

So how does this enable us to understand changes in telecommunications regulation? From this all-to-brief and rather crude account of Gramsci's perception of the symbiotic relationship between capitalism and the superstructural elements of political and civil society we can extract an awareness that changes in the nature of production are reflected in regulatory frameworks. These frameworks, in turn, facilitate the completion of the movement from one industrial order to another, as well as assist in the globalisation of the capitalist mode of production. Put differently, regulatory frameworks are, at one and the same time, the result of, and agents for change. For our purposes, we can hypothesise that the growing significance of telecommunications will necessarily be reflected in the development of transnational and global regulatory frameworks. In turn, these frameworks will facilitate the further expansion of the telecommunications sector. Moreover, as production in this sector further globalises, so too will the legal apparatus for smoothing its advance; and, as this process of globalisation accelerates, so will the pressure for the creation of a central regulatory authority. With this in mind we can begin to examine recent developments in telecommunications regulation.

**Telecommunications: an historical perspective**
Historically, the development, operation and control of telecommunications has been a state-centred activity. The provision of a largely standard system across which voice signals were transmitted was supported by the (complementary) twin pillars of natural monopoly and universal service. Broadly-speaking, the desired outcome of facilitating as many users as possible with a basic telephone service at a uniform price could be most efficiently achieved, it was argued, through either state-owned (for example, in Europe and Japan) or private, though highly regulated, (for instance, the USA) monopoly, the result of the high start-up and roll-out costs associated with providing a fixed-link telecommunications network. Telecommunications equipment producers and telecommunications service providers, certainly in northern hemisphere markets, were almost exclusively domestically-based. The provision of international telecommunications services was an intergovernmentalist affair which aimed to ensure the development of enough technical standardisation to make the system operable and to create an acceptable management structure for scarce airwave resources (Cable and Distler, 1995: 5). Service provision across states was, then, merely a question of co-ordination rather than regulation.

However, in the closing decades of the twentieth century a well-documented series of factors contributed to the mutation of telecommunications from a nationally-focused to an international, if not truly global, sector (Hills, 1986; Noam, 1992; Steinfield, Bauer and Caby, 1994). Key developments in technology, such as the digitisation of telecommunications switches and the emergence of fibre-optic cable, greatly increased the research and development burden faced by producer firms, prompting them to seek a larger, inevitably international, market base to shoulder the cost. The application of computer technology to telecommunications created not only the possibility of a new multimedia services paradigm, but also pressure from computing firms to break into the closed-shop of telecommunications service provision, and from frustrated telecommunications users from the powerful multinational business community for them to be allowed to do so. These user organisations campaigned for the creation of high speed, sophisticated networks with enough global reach to facilitate their international expansion. In response to this new lucrative market, the sleeping giants of, first, telecommunications equipment production, then, service provision, engaged in a series of international forays into competitor markets, resulting in a host of strategic alliances and take-overs. It is here that we see the emergence of pressures for a transnational regulatory framework to facilitate their expansion.

None of the globalisation of telecommunications business witnessed in recent years could have occurred without a sea-change in the regulatory philosophy and arrangements which existed in nation-states. The complex process of loosening the regulatory strictures which historically governed telecommunications markets was underpinned by the emergence of neo-liberal economic arguments stressing the merits of liberalisation and free competition. This was witnessed most significantly in the US in the 1980s, where AT&T was divested of its regional telecommunications companies, forced to concentrate on the carriage of long distance telecommunications traffic and, later, permitted to enter the computer services market (Braithwaite and Drahos, 2000: 325). In Europe, the UK government of the 1980s, an ideological bedfellow of the Reagan administration in the US, was next to follow suit (see Overbeek, 1990). Likewise, as a result of domestic user and US government pressure, in 1984 Japan partially privatised its telecommunications service provider, NTT, opening up its markets to (limited) competition (Cable and Distler, 1995: 12). Since then, nearly all governments in Europe, Asia and North America have pursued a liberalisation agenda.

The seemingly inexorable liberalisation and globalisation of telecommunications has created two important developments in telecommunications regulation. First, international trade and investment, historically not an issue of major significance in telecommunications due to closed and protected national markets with chauvinistic public procurement policies, has assumed vital importance. Second, as it became readily apparent that national regulatory structures and existing international arrangements would be incapable of catering for the new era of liberalisation in telecommunications, thoughts turned to the creation of transnational regulatory structures at the regional and ultimately global levels for this purpose.

**Regional Regulation**
It is logical, then, that we should see the development of transnational regulatory apparatuses occurring first at the regional level, established on the principle of geographical familiarity and the relative cost advantages of spatial proximity. The EU provides an important example of such developments in telecommunications. It was not until the early 1980s that telecommunications began to become a significant policy domain for the EU. However, since then there has been a progressive re-regulation of many of the most important aspects of the sector from the national to the European level, with a consequent cedence of regulatory control by nation-states. The origins of the process itself were the consequence of the complex interaction between a number of important economic, political and institutional factors which are reflected in our neo-Gramscian understanding of regulatory change.

The already highlighted key technological changes in telecommunications and economic and commercial opportunities which they afforded were seized upon by a miscellany of corporate interests and institutional actors to propound arguments for the transnationalisation of key aspects of telecommunications regulation to the EU level. Neo-liberal oriented national governments (most prominently the UK), the European Commission, as well as corporate business interests from the information technology and telecommunications business user communities, pressed hard for change. A curious mix of Euro-centrist industrial policy and economic liberalism arguments were put forward to support a redrawing of the contours of the telecommunications sector in Europe.

The European Commission acted as an important political advocate and facilitator for the liberalisation agenda of business, in return for which it received very significant institutional aggrandisement. The Commission played a key role in underlining the direct and secondary strategic economic and industrial importance of telecommunications, embellished with dire warnings of the dangers of lagging behind Europe's main competitors: Japan and the US (European Commission, 1984). The only appropriate course of action was deemed to be the liberalisation of telecommunications markets, a call which became couched in the wider political-economic project of creating a Single European Market (European Commission, 1985) – the two projects were undoubtedly mutually reinforcing (European Commission, 1986: 5). In this respect, the roots of a 'received wisdom' in relation to telecommunications were developed which gradually gained ground across the EU in the late 1980s and 1990s. A vital element in the solidification of any new transnational context is the creation of a set of legally binding arrangements in which a system – and, in extension, particular modes of behaviour – becomes embedded. This was soon to follow.

However, it was also clear that there was significant and powerful opposition to the creation of such a new regulatory system. This opposition came from a complex of traditional telecommunications interests, comprising national member states ideologically opposed to liberalisation and the loss of national control of key industries, economically weak EU member states, and monopoly public telecommunications service operators (PTOs). This tension was reflected in the 1987 Green Paper proposed by the European Commission (European Commission, 1987) which aimed to push re-regulation forward through advocating the EU-wide liberalisation of telecommunications terminal equipment and Value Added Network Services (VANS), while guaranteeing member states rights to keep the structure of provision of voice telephony as a publicly-owned monopoly, if so desired. Subsequently proposed liberalisation measures (directives) in these areas (European Commission, 1988; European Commission, 1989) brought the conflict to a head, illustrating the determination of the European Commission – backed by corporate business users and tacit support from powerful member states such as the UK and Germany – to drive forward the liberalisation agenda. Aware of the national member state veto on proposed legislation which existed at the time in the EU, the Commission took the tenacious step of introducing both the terminal equipment and VANS directives into law without securing the approval of the EU Council of Ministers, a move which survived two legal challenges in the European Court of Justice from the more dirigiste member states and required a political compromise to be stitched together in respect of the services issue (Schmidt, 1998: 173).[1]

The power of multinational business interests in shaping both the agenda of the Commission (see Esser and Noppe, 1996; Levy, 1997) and the preferences of national governments in telecommunications cannot be over-stated. It was argued that telecommunications, whose costs parameters (in terms of investment in new technology) required markets much larger than those available in any national jurisdiction, needed to be reconfigured into an international arena. However, such a transition contained within it huge risks for the nation-state and its established strategic industrial interests in telecommunications. Thus, as our conceptual framework suggests, despite the significant cultural and linguistic differences in Europe, the EU was chosen as a focal point for this experimental exercise in the regionalisation of telecommunications. As the 1990s began, an important addition to the adherents of the transnationalisation-liberalisation agenda emerged in the shape of certain large telecommunications service providers. These organisations, only a few years previously reticent about liberalisation for fear of loss of market share to more competitively fit players from elsewhere in Europe and the US, gradually became more confident of succeeding in a liberalised international telecommunications market and (thus) advocates of a set of regulations for its governance. The more wary EU member states (such as France) also became somewhat reluctant acceptors of the argument that national self-interest (in terms of commercial performance, job creation and economic growth) could be best pursued in a more internationalised telecommunications sector, even though such a scenario contained new risks and uncertainties.

The emerging 'consensus' on telecommunications in the EU was legally cemented in two Resolutions by all member states agreeing to liberalise voice telephony services and infrastructure by 1998 (European Council of Ministers 1993; 1994). This precipitated, by necessity, the passage of a swathe of legal measures forming the regulatory parameters of the new system and representing an unprecedented transnationalisation of the European telecommunications regulatory regime (see European Commission 1994, 1995, 1996a, 1996b; European Parliament and European Council of Ministers 1997a, 1997b). Since 1998,[2] the liberalised EU telecommunications system has been operationalised and further refined. A major review of regulation in the communications sector occurred in 1999 (European Commission, 1999), significantly influenced by the debate launched by the Commission in 1997 on the possibility of creating a converged European level regulatory structure to cater for all information and communications infrastructure and content (European Commission, 1997). As a result, a revised more convergent regulatory framework has been proposed to cater for all telecommunications infrastructures and associated services provision, though, importantly, the new framework will not for the foreseeable future cover broadcasting services content or electronic commerce (European Commission, 2000a). The proposal, if agreed by member states, will be built around a series of six directives on horizontal regulatory arrangements, authorisations, access and interconnection, universal service, data protection and privacy and local loop unbundling.[3]

---

[1] The Commission legitimised its action on the basis of Article 90(3) of the Treaty of Rome which instructs it to take rectificatory action to get rid of abuses of dominant position by public undertakings in markets where it is found to exist. In the Commission's view, telecommunications terminal equipment and VANS were examples of this.

The EU thus provides a useful case study of the development of a transnational regulatory regime, based on neo-liberal principles and practices. Crucial to the system's functioning is the relationship between the national and European level. While legal powers of enforcement of agreed legislation rests at EU level, ensuring that the system works effectively in practice requires collaboration between the European Commission in its role as guardian of the treaties establishing the EU, the European Court of Justice with legal powers to force compliance with legislation and, finally, the series of independent National Regulatory Authorities (NRAs) established to regulate each national market. Here, evidence exists of problems related to a number of matters of transposition and implementation of EU legislation, such as licensing procedures, cost accounting systems, carrier pre-selection, access and interconnection, competitive advantages enjoyed by the incumbent operator and monitoring of consumer issues nationally (European Commission, 2000b). There have been calls, particularly from newer telecommunications corporate interests, for the establishment of a European regulatory authority (Bartle, 1999). That said, the current proposal for a European level High-Level Communications Group, while in theory independent, will only operate in an advisory capacity and comprise members designated by the NRAs.

---

[2] Luxembourg (2000), Greece, Spain, Portugal and Ireland (all 2003) were given compliance derogations from the 1998 agreement
[3] The directive on local loop unbundling was agreed in December 2000. It is envisaged that the remaining proposed directives in the new framework will be ratified simultaneously (European Commission 2000a)

It is important to note that the emergence of the EU regulatory apparatus effectively bypassed the old co-ordinative regime of the CEPT, a pattern also replicated at the global level with the shift of power from the International Telecommunications Union (ITU) to the WTO (to which we return below). The EU, as a transnational focal point for regulation, proved attractive not only for economic and spatial reasons, but in terms of exercising political control. The well-established, though often turbulent, political-economic project of European integration was a useful institutional context within which to propound and establish a new agenda. It allowed national member states to operate within familiar institutional territory in which policy issues are often packaged together for bargaining purposes and where the ultimate right of veto existed until recently.

*Global policy issues in the EU telecommunications*

From the outset, the development and adoption of a concerted EU position in international telecommunications fora was ensconced as a priority area (European Commission, 1984), with the European Commission assuming the important political role of EU representative in the process (Mansell, Morgan and Webber, 1989). An early example of the attempt to establish the new competitive agenda of telecommunications concerns the US government and European Commission's efforts to drive reform of the system of international call charging between member countries of the ITU. There is clear evidence that in doing this, both parties championed the agenda of multinational telecommunications business users as well as liberalised international telecommunications carriers, in the case of the US, which was not a member of the system. The US attempted to reduce the ITU's regulatory power in the international domain by advocating that such issues should be the preserve of the General Agreement on Tariffs and Trade (GATT).

In the early 1990s, the European Commission vociferously complained against Telecommunications Operator (TO) resistance towards the growth of privately operated telecommunications networks for business users and went as far as to intervene in the ITU's International Telegraph and Telephone Consultative Committee (ITTCC) D-series of negotiations, by threatening to require TOs to open up their accounting systems to scrutiny. The Commission claimed, with considerable justification, that users were being overcharged by telecommunications service providers by as much as $20 billion per year (*Financial Times*, 29 January 1991). Since then, international call charges have fallen dramatically (European Commission 2000c[4]). Nonetheless, the US government has continued to promote the interests of its multinational companies claiming that a considerable trade deficit is accrued annually in international telecommunications traffic, due to artificially high rates charged by competitor countries.

---

[4] For example, in the 2000, the average price of international calls decreased by 15.1% in Europe and the price of leased lines for the carriage of national and international traffic fell by 30% between 1997-2000.

However, it has been suggested that, due to the use of the 'call-back' system (which ensures that calls can be routed through locations which have low international call charges), while the US paid out US$5.7 billion to other countries in 1996, in the same year its telecommunications carriers received nearly US$14 billion. In development terms, it has also been argued that a net transfer of resources from the northern to the southern hemisphere as a result of accounting rate differences is a positive thing (Cane, 1996).

In recent years, the EU regional telecommunications bloc has played an important part in global trade negotiations taking place in the WTO, where a landmark agreement was reached among 69 signatories (now 84) in February 1997 to liberalise markets in basic telecommunications services (the Agreement on Basic Telecommunications). In line with its economic self-interests, the EU has attempted to impose its policy model on the global system, arguing for the extension of the agreement to cover more advanced telecommunications infrastructure services related to the carriage of electronic mail, online information transaction processing and electronic data interchange (EDI) (European Commission 2000c: 2). It has been forcefully argued that 'the EU has very strong *offensive* interests in services, and we must push them forward' (Lamy, 2000: 2), of which telecommunications represents a significant part. It is important to note that EU proposals, put forward as part of the next round of General Agreement on Trade and Services (GATS) negotiations would exclude liberalisation of content creation and provision, reflective of the recently agreed new communications regulatory framework at EU level. In the process, the Europeans have also expressed dissatisfaction with the interpretation and implementation of the 1997 Basic Telecommunications Agreement on issues such as exemptions, deployed by some signatories, to the Most Favoured Nation (MFN) principle; flexible interpretation of what is covered by the agreement; market access arrangements; prolonged phasing in of the Agreement's measures; and market regulation issues, such as licensing, interconnection, universal service and the like (European Commission 2000c: 2-3).

In its deliberations on global telecommunications trade systems, there is evidence of the EU working hard to ensure a consensus on the benefits of the liberalisation agenda. The Commission's Directorate-General responsible for trade has argued that its newly extended proposals will provide many social and economic benefits to developing economies (European Commission, 2000d). Liberalising trade in services and creating a predictable regulatory environment for their delivery will facilitate the attraction of foreign direct investment and improve the infrastructures for financial services, transport and telecommunications (European Commission 2001a). It could also be argued, however, that if not carefully monitored by national governments, multinational business interests will benefit more from these events than the indigenous economy. In its proposals to the current round of GATS negotiations, the European Commission has declared the principles of the telecommunications model an exemplar of how agreements should be formed in other contexts (European Commission, 2001b). It has also been argued that it is in the interests of the EU and the US to persuade those opposed to globalisation, both within their territories and the developing world, of its benefits (Lamy 2001: 2-4).

**Global Regulation**

Like the US, then, the EU has an interest in not only moving the regulatory agenda forward through a consolidation of global telecommunications regulation, but also in ensuring that such regulation better reflects the needs of European telecommunications concerns. However, pointing to the development of transnational regional regulatory frameworks – exemplified most prominently, though not exclusively, by the EU – and the energy for their further development, offers only a snap-shot of developments in global telecommunications regulation. To understand more comprehensively the full extent to which a system of telecommunications regulation has emerged we need to explore something of the development of a wider global regulatory structure.

Reflecting the changed exigencies of contemporary capitalism, the creation of the WTO on 1 January 1995 brought with it a deepening and widening of the arena of commercial activity subject to regulation. The WTO's legal framework is designed to systematically erode exceptions to liberalisation by drawing previous exempt areas under GATT rules (the GATT is now one of a series of commercial agreements administered by the WTO) – most notoriously textiles and clothing, and agriculture – and limit the period in which members can seek exemptions from MFN. But perhaps the most significant inclusion has been the extension of trade regulation to include trade in services (under the GATS), of which telecommunications is but one, albeit a crucial, dimension.

Reflecting the changing complexion of industrial economies is not, however, the end of the story. The WTO's creation also brought with it a widening of the arena of trade regulation to include the Agreement on Trade-Related Aspects of Intellectual Property Rights (TRIPs) and the Agreement on Trade Related Investment Measures (TRIMs). These agreements have taken trade regulation beyond its traditional parameters to include areas deemed to be 'trade-related' – areas which, though not necessarily relating to tradable entities, are deemed intrinsic to the production process. Both the TRIPs and the TRIMs have strategic importance for this emerging system of global regulation. Neither agreement seeks to bring about a direct expansion in commercial activity. Rather, their aim is to further smooth the transnationalisation process. The TRIPs, for instance, begins the process of securing the global ownership of intellectual property; while the TRIMs goes some way towards the liberalisation of investment flows (see Wilkinson, 2000: 56-68). Crucially, and in a change of practice to the GATT-administered system, as part of the Uruguay Round accords members were required to ratify the full complement of regulations contained within the WTO's legal framework as part of a 'single undertaking'. Exceptions to any of the 29 agreements administered by the WTO at the time of accession were not permitted,[5] thus cementing the regulatory significance of the WTO as well as consolidating the liberalisation process.

Yet pointing to the deepening and widening of the arena of trade regulation does not uncover the full significance of the WTO's establishment. Also of key importance to the WTO's legal framework are a series of provisions which outline the Organisation's role in the development of a coherent global system of regulation. These provisions locate the WTO squarely at the heart of a system of regulation comprising the International Monetary Fund (IMF) and World Bank and a host of other organisations, of which the World Intellectual Property Organisation (WIPO – a partner organisation through the TRIPs) is perhaps the most significant. These provisions not only restate the mutually harmonious interrelatedness of the work of each organisation, but also remove the potential for each body to operate in contradiction to the others. This is the removal of so-called 'cross-conditionality' (see the *Declaration on the Contribution of the World Trade Organisation to Achieving Greater Coherence in Global Economic Policy Making* annexed to GATT, 1994). Furthermore, in the case of the WIPO, they locate at the heart of that system the principle of global ownership of intellectual property rights.

Two further developments reveal the extent to which a single regulatory system has begun to crystallise. First, the WTO's legal framework requires that *all* regional free trade areas and customs unions in which member states (such as the EU, the North American Free Trade Agreement (NAFTA) and Asia Pacific Economic Co-operation (APEC) among others) participate register with the Organisation and conform to its body of rules[6] – in doing so, increasing the pressure for convergence around a single regulatory model. Second, the convergence around a single model can also be found in the way in which other, non-WTO agreements, have begun to adopt the format of WTO agreements. The much lambasted Multilateral Agreement on Investment (MAI), for instance, was designed around those core principles – MFN, national treatment and reciprocity – that form the spine of the GATS, GATT, TRIMs and TRIPs (Wilkinson, 1999: 181; 2000: 43-51).

---

[5] That said, members can request exceptions to the GATS by lodging them on the so-called 'negative list' during accession negotiations. Further exemptions can only be made by requesting an MFN waiver from the Ministerial Conference.

[6] See Article XXIV of the GATT.

*The WTO and Telecommunications*

It is as part of this wider framework that we find moves to develop a global system of telecommunications regulation. The completion of the Uruguay Round witnessed a series of limited commitments to liberalise telecommunications contained in the WTO's legal framework, under the auspices of two annexes to the GATS and a Ministerial Decision on Negotiations on Basic Telecommunications. These provisions were intended to begin the process of liberalising the telecommunications sector across the Organisation's members, and set out a framework for future negotiations directed at accelerating the liberalisation process.

Two features of the WTO's provisions on telecommunications require elaboration. The first is institutional and relates to the nesting of institutions centred around the WTO in the development of a system of global regulation. Like the TRIPs, the GATS makes provision for the development of relations with other organisations in instances when to do so is of benefit to the overall objectives of the WTO (Article XXVI of the GATS). This provision is modified by paragraphs 7 (a) and (b) of the GATS annex on Telecommunications. Here reference is made to the work of others organisations both inter- and non-governmental. Of these, only two are mentioned by name – the ITU and the International Organisation of Standards (IOS) – albeit that the ITU is given greater emphasis. The provisions do not specify the nature of any future relationship between the organisations; rather, they open up space for the future development of a meaningful relationship by recognising the importance of the ITU and IOS in nurturing an international standards regime and the potential for such a regime to be utilised by the WTO. The potential exists, then, for the ITU and the IOS to be drawn into an institutional complex wherein their prior work and expertise is utilised as the platform upon which a process of standardisation is initiated.

Second, the annex to the GATS also contains a commitment to take further liberalisation in telecommunications by establishing the Negotiating Group on Basic Telecommunications (NGBT – since replaced by the Group on Basic Telecommunications (GBT)). Part of this commitment involved the establishment of a surveillance capacity for the NGBT designed to ensure that in the period between the completion of the Uruguay Round accords and the completion of negotiations on basic telecommunications, member states did not implement measures enabling them to develop or increase an advantage during the negotiations (such as by imposing a more discriminatory system which would then be the subject of negotiation).[7] But these surveillance activities also ensure that the process of standardisation is given additional impetus as the telecommunications regimes of nation-states are opened up to minute scrutiny. As is well-known, these provisions laid the foundations for the conclusion of the Agreement on Basic Telecommunications on 15 February 1997, with an entry into force date of 5 February 1998 (see Blouin, 2000).

---

[7] Paragraph 7 of the Ministerial Decision on Basic Telecommunications.

The effect of the Agreement on Basic Telecommunications, albeit for some modest in the extent of its liberalisation, has been to deepen the extent to which telecommunications regulation is incorporated in the WTO's regulatory system. Coupled with the potential for meaningful co-operation to develop between the WTO and the ITU and IOS, recent moves have embedded the telecommunications regulation in a global framework and represent a significant development in the general movement of regulatory sovereignty away from national authorities, through regional conduits, to a global system. Moreover, the mechanisms put in place for the further liberalisation and standardisation of this sector will consolidation this movement further.

The WTO's significance, then, lies in its organic structure. This structure is created such that it draws together a range of organisations into a constellation of bodies, each working in concert to promote a particular model of regulation. This system comprises not only global bodies such as the IMF, World Bank, WIPO, ITU and IOS, but also the regional projects underway in Europe, North America, the Asia-Pacific and elsewhere. For us, this movement represents a development of the superstructural elements of political authority necessary to assist in the mutation of capitalism. Further weight is given to this when we look at the content of the system of regulation administered by the WTO. We see in this framework a reflection of the changed exigencies of contemporary capitalism. The inclusion of the GATS and the Agreement on Basic Telecommunications, as well as the incorporation of the TRIPs and TRIMs provide ample illustration. In extension, it is reasonable to suppose that further developments in this system of regulation will seek to meet the regulatory requirements of ICT-based capitalism, among others. Indeed, much of the tension that currently afflicts the WTO centres around a push by industrial states for a move into global regulation on government procurement, investment, information technology, biotechnology, telecommunications, and services, and resistance to such moves by developing countries whose interests lie in securing the implementation of existing agreements and issues of special and differential treatment (Wilkinson, 2001).

**Conclusion**

In this paper we argue that insight into regulatory development and change can be gleaned by utilising something of Gramsci's concepts. We see the development of regulation as a necessary reflection of the changing needs of capitalism. There are two dimensions to this. First, as pressures to move beyond national markets build, so too will those for the development of a supporting regulatory framework. Second, the content of this regulation will reflect the needs of transnational capitalism and, in extension, smooth its advance.

Such a framework enables us to offer an interpretation of recent changes in telecommunications by locating them within wider global developments. What we see emerging is a complex multilayered system wherein regulatory sovereignty is passing from the nation-state through regional bodies to global organisations. This regulatory movement is, however, incomplete. The WTO has begun to set out the parameters of a global telecommunications framework, but significant obstacles remain. These obstacles lie not only in the current tension between developing countries and their industrial counterparts, but also in the nature of ICT-based capitalism, which has yet to be fully established.

The consequences of the development of a new, more convergent system of telecommunications regulation are only beginning to emerge. For instance, the possible extension of the WTO's remit to include more content-rich communications services, already being propounded by the EU and US, will herald intense negotiations and may be difficult to achieve. Regulatory approaches to these value-added elements of communications services content differ markedly, even among the most enthusiastic neo-liberal states. This of course says nothing of the lack of an infrastructural capacity in developing states (Schiller, 1999). Here, a lot will hinge on the realisation of infrastructural and economic welfare gains promised so confidently and vociferously by those advocating liberalisation. That said, it remains the case that the development of a regulatory framework favours those lead industrial states with a burgeoning interest in defining the parameters of telecommunications regulation. It does not favour those developing states whose principal economic interests lie in seeking renewed ways in which to improve the well-being of their populations and who will, inevitably, be left behind in the race for market position in the digital era.